# Presenting the SWTC:
# A Symbolic Corpus of Themes from John Williams' *Star Wars* Episodes I-IX

CLAIRE ARTHUR[1]
*Georgia Institute of Technology*

FRANK LEHMAN
*Tufts University*

JOHN MCNAMARA
*Georgia Institute of Technology*

ABSTRACT: This paper presents a new symbolic corpus of musical themes from the complete Star Wars trilogies (Episodes I-IX) by John Williams. The corpus files are made available in multiple formats (.krn, .sib., and .musicxml) and include melodic, harmonic, and formal information. The Star Wars Thematic Corpus (SWTC) contains a total of 64 distinctive, recurring, and symbolically meaningful themes and motifs, commonly referred to as *leitmotifs*. Through this corpus we also introduce a new humdrum standard for non-functional harmony encodings, `**harte`, based on Harte (2005, 2010). This report details the motivation, describes the transcription and encoding processes, and provides some brief summary statistics. While relatively small in scale, the SWTC represents a unified collection from one of the most prolific and influential composers of the 20th and 21st century, and the under-studied subset of film and multimedia musical material in general. We hope the SWTC will provide insights into John Williams' compositional style, as well as prove useful in comparisons against other thematic corpora from film and beyond.

AS the study of large musical corpora via techniques of digital encoding and analysis continues to grow, one promising yet relatively unexamined repertoire is that of film music. Original scores written for movies represent an expansive and culturally pervasive source of music, one lying at the intersection of two idioms—the European extended common practice and 20th-21st century popular genres—the latter of which has received considerably more attention in the corpus studies literature. The digital encoding of film music is particularly challenging for a number of reasons: stylistic variability, ambiguity as to what counts as a typical or salient representation of a given musical idea, instability or absence of a definitive notated text, and use of parameters less easily represented by current encoding systems (i.e., timbre, audiovisual synchronization, musical topics). While none of these issues on their own are unique to this corpus, their conjunction, and in some cases extremity, is a defining feature. Yet, these same challenges also are part of the reason why the repertoire is appealing for corpus study, as they force one to refine tools and expand what can be done through digital encoding. In this data report, we present the *Star Wars Thematic Corpus* (SWTC): a digital encoding of the thematic material from the Original, Prequel, and Sequel *Star Wars* trilogies by John Williams.

## MOTIVATION

*Star Wars* is one of the biggest film franchises of all time (McKinley, 2019). The importance of music—and particularly iconic and recurring themes—have often been discussed as being a key part of the series's lasting success and wide appeal (Golding 2019). John Williams's contributions to the nine live-action films that make up the "Skywalker Saga" and its interlinked Original (1977-1983), Prequel (1999-2005), and Sequel (2015-2019) trilogies have had a significant and continuing impact on cinematic scoring since the late 1970s (Scheuer 1997, Cooke 2008, Audissino 2021). As an object of study, the music from the *Star Wars* films is particularly useful as a vehicle for corpus analysis given that the music was written by a single composer across all major series entries and over a period of four decades. These scores represent a high degree of compositional craft and an overall unified symphonic style. At the same time, given the gradual development of Williams's style and changing film-scoring trends since the 1970s, this corpus offers a basis or comparison against itself, as well as preexisting Classical and pop/rock copera and future film-musical corpora.

## THE STAR WARS THEMATIC CORPUS

### Access

All files in the Star Wars Thematic Corpus (SWTC) are downloadable via Github repository (https://github.com/Computational-Cognitive-Musicology-Lab/Star-Wars-Thematic-Corpus ) as a publicly accessible symbolic (human- and machine-readable) corpus of film music, and is open source under a Creative Commons license (CC-BY-NC 4.0, https://creativecommons.org/licenses/by-nc/4.0/ ). We also include a spreadsheet in the repository that contains additional metadata not referenced in the humdrum or musicxml files, such as the number of appearances of the theme throughout the original full score. As we plan to update this corpus in the future, we release the current corpus (detailed here), released in 2022 as version 1.0.

### Description

The nine *Star Wars* scores each contain between two and three hours of originally composed music. The percentage of score material from recurring themes is between roughly 20% (Episode II) and 40% (Episode VIII) (Lehman 2018a). Out of this large oeuvre, there can be discerned 64 distinctive, recurring, and symbolically meaningful themes and motifs, commonly referred to as *leitmotifs*. While there exist no authoritative guidelines for what should and should not be designated a leitmotif, the classification and naming of these materials is based on rigorous criteria developed in Bribitzer-Stull (2015) and systematized in Lehman (2018a, 2022). The two key criteria for inclusion are: 1) musical distinctiveness and 2) recurrence across at least three discrete spans of film music (cues). The relative openness of this definition, particularly the distinctiveness condition, permits certain themes that might otherwise go unnoticed to be admitted into the corpus (ex. "Revelations," "Happy Landing," "Map"). Excluded at this time are alternative forms of musical materials (variations), musically indistinct motives (incidental themes), memorable but non-repeating materials (set-piece themes), source music, and music composed by anyone other than John Williams.

Leitmotifs in *Star Wars* are also highly variable in their appearances, and may lack the sorts of clear or definitive statements typically sought in corpus studies. For the purposes of the SWTC, a judgment call was made as to the most representative and/or characteristic guise of a given theme, sometimes involving a degree of idealization from any one specific usage (e.g. "Leia's Theme"). While some themes are internally complex, multi-sectional, or structurally variably, only the most basic and distinguishing measures of a leitmotif were included. Future expansions of this corpus may reintegrate these materials, much of which have already been transcribed and encoded.

Per typical procedures for encoding musical corpora, a reductive approach has been taken for all 64 themes, with a single melodic line with only one note at a time represented and harmonic information conveyed through chord symbols and harmonic function with Roman numerals in cases where a tonal center is clear. Symbolic indicators for keys, cadences, and four kinds of formal boundary (full, sectional, abrupt, and half-cadential conclusions). All transcription work was performed by author F.L..

At times, the limits of what chord symbols and Roman Numerals can convey are pressed. Williams's harmonic language is quite sophisticated (Lehman 2018b), informed by his background in jazz (as discernible in themes "Luke and Leia" and "Gloomy Courtship") and the growing popularity of minimalist techniques like ostinati, drones, and static harmony (as operative in "Duel of the Fates," "Snoke," etc.). Standard chord symbols do a poor job conveying the active and harmonically-generative basslines that undergird many leitmotifs like the "Main Theme" or "TIE-Fighter Attack." In such cases, the SWTC employs a combination of liberally included extensions and suspensions, slash and polychord notation, and a dedicated pedal point data layer for dissonant bass notes. Importantly, these pedal indicators do not correspond to fixed bass pitches per se, but rather moments when a given triadic chord is not supported by a consonant chord-tone in the bass.

**Digital Encoding and Representation**

All 64 files in the SWTC are available in multiple formats: their original Sibelius format, uncompressed musicXML format, and humdrum format (Huron, 1995; Sapp, 2021). In the humdrum files, the musical information is preserved in time-aligned columns, called spines, where each spine contains a different type of musical information. (For full humdrum documentation, see https://www.humdrum.org/guide/). We include the following spine types in the SWTC humdrum files: **kern, **harte, **harm, **altharm, **pedal, **cadence, and **text. Both **kern and **harm are preexisting humdrum format standards (see https://www.humdrum.org/guide/), whereas the rest have been customized for this specific corpus. For the SWTC, we propose a new standard humdrum representation for chord symbols, **harte, based on an existing proposed standard for representing chord symbols (Harte, 2010) widely in use already within the music information retrieval community.[1] Descriptions of each kern spine can be found in Table 1 below. Since not every file contains information for the relevant spine, the number of spines per file is variable. At a minimum, each file must contain a **kern spine (melodic information) , and may optionally include all other spines as relevant. For instance, themes with no clear tonal center do not contain **harm or **altharm spines (functional harmony representations). Themes with explicit harmonic accompaniment will include a **harte spine (and **harm spine if appropriate), however, files with a **harm spine but no **harte spine are indicative of implied harmony only. The **altharm spine allows for alternative plausible functional-harmonic interpretations for any single chord event—a fairly common occurrence in Williams's enriched tonal vocabulary. (For example: the penultimate chord in "Yoda's Theme" is analyzed a first as V[b9], but contextual features justify an alternative, more-plagally oriented interpretation of iv[add6] over a G-pedal.)

**Table 1.** Kern spines (columns), and the description of each, for the SWTC.

| | |
|---|---|
| **kern | Melodic information in humdrum's standard kern format. See https://www.humdrum.org/guide/ch02/ |
| **harte | Chord labels based on Harte, 2010. A brief explanation can be found below and a more extensive explanation in the documentation on the SWTC github repository. (See: https://github.com/Computational-Cognitive-Musicology-Lab/Star-Wars-Thematic-Corpus). |
| **harm | Roman numerals according to the **harm format. See https://www.humdrum.org/rep/harm/ |
| **altharm | Alternative Roman numeral interpretations in **harm format. If none, then the spine is empty with only placeholder tokens. |
| **pedal | Indication of pedal tones. Pedal tones are assumed to continue unless explicitly overtaken by a new pedal tone or replaced with "None" |
| **cadence | Cadential information and Formal boundary markings (e.g., ";;" signifies an abrupt or non-conclusive ending). Full list of cadence types include: Perfect Authentic (PAC), Imperfect |

---

[1] For example, the Harte standard has been adopted for use in the MIREx competition's audio chord detection task since 2009 (see: https://www.music-ir.org/mirex/wiki/MIREX_HOME). This has led to several datasets adopting this standard for annotation (e.g., McGill Billboard Corpus: Burgoyne et al. (2011); Guitarset: Xi et al. (2018)). Further anecdotal evidence can be found in McFee & Bello, 2017.

|  | |
|---|---|
|  | Authentic (IAC), Half (HC), Plagal Half (PHC), Chromatic (CC), Chromatic Half (CHC), Modal Half (MHC), and Evaded (EC). Boundary markings are encoded as follows: Conclusive ending `]]`, Half-cadential-like ending `//`, Sectional ending `\|\|`, and Abrupt or non-conclusive ending `;;`. |
| `**text` | Any additional information or notes in comment-style form. E.g., type of modulation (e.g., pivot), textural notes, etc. |

Each file includes relevant bibliographic metadata in the humdrum reference records (see: https://www.humdrum.org/reference-records/). Reference records at the top of each file include: the composer, title, associated work, date of composition, transcriber, and digital encoders. The conversion of the corpus into humdrum format, including encoding details and decisions, was shared by authors CA and JW.    . An example excerpt from one file is shown in Figure 1.

```
!!!COM: John Williams
!!!OTL: The Imperial March (Darth Vader's Theme)
!!!GAW: Star Wars: Episode V: The Empire Strikes Back
!!!ODT: 1980
!!!OCL: Frank Lehman
!!!ENC: Claire Arthur
!!!ENC2: John McNamara
**kern      **harte     **harm      **altharm   **pedal     **cadence
*clefG2     *           *           *           *           *
*k[b-e-]    *k[b-e-]    *k[b-e-]    *           *           *
*g:         *g:         *g:         *           *           *
*M4/4       *M4/4       *M4/4       *           *           *
*MM110      *MM110      *MM110      *           *           *
=1-         =1-         =1-         =1-         =1-         =1-
4g          G:min       i           .           Pedal: G    .
4g          .           .           .           .           .
4g          .           .           .           .           .
8.e-L       E-:min      vi          .           .           .
16b-Jk      .           .           .           .           .
=2          =2          =2          =2          =2          =2
4g          G:min       i           .           .           .
8.e-L       E-:min      vi          .           .           .
16b-Jk      .           .           .           .           .
2g          G:min       i           .           .           .
=3          =3          =3          =3          =3          =3
4dd         G:min       .           .           .           .
4dd         .           .           .           .           .
4dd         .           .           .           .           .
8.ee-L      E-:min      vi          .           .           .
16b-Jk      .           .           .           .           .
```

**Figure 1.** Sample encoding of a selected file from the SWTC (showing measures 1-3 only), "Imperial March" theme (file #10 from the 'Original Trilogy' corpus).

## Harte Encoding

Trevor DeClercq (2015) discusses a well-known issue in the field of computational musicology relating to the lack of any standardized encoding format for harmony. He comments that *"each format has its advantages and disadvantages", and that while "it is not impossible to convert and translate between different encoding formats…it makes this type of research much more difficult."* The introduction here of a new non-functional harmony encoding in humdrum format is aimed to facilitate data sharing, and provide a more robust encoding method for music containing harmony where traditional Roman numeral interpretations frequently become inappropriate or inconsistent. We chose the Harte (2005, 2010) representation since it provides a comprehensive method for encoding any chord as well as a simple, condensed method for common chords, and because it is already widely encountered in the music information retrieval (MIR) community (see footnote 1).. As illustrated in Harte's basic encoding scheme shown in Table 2 (from Harte, 2005, p.69), the most common chord types (triads and seventh chords) can be represented

*Empirical Musicology Review*          Vol. X, No. X, 20XXwith a "Root: shorthand" notation (column 2), which contain the root plus the intervals above the root as shown in column 3. For example, a C major chord would be written as "C:maj", a G major-minor seventh chord as "G:7", or an F chord with a suspended fourth as "F:sus4". Omissions can be indicated in the interval list with an asterisk (e.g., no third = *3), and additions in parentheses. In addition, non-common chords can always be represented by indicating any root note (even if arbitrary) and the collection of intervals above it (e.g., "C:(3,#5,6,7)"). See Harte (2010) for a complete explanation.

**Table 2:** Shorthand Descriptions of Common Chords (Harte, 2005)

| Chord Type | Shorthand Notation | Interval List |
|---|---|---|
| **Triad Chords:** | | |
| Major | | (1, 3, 5) |
| Minor | min | (1, b3, 5) |
| Diminished | dim | (1, b3, b5) |
| Augmented | aug | (1, 3, #5) |
| **Seventh Chords:** | | |
| Major Seventh | maj7 | (1, 3, 5, 7) |
| Minor Seventh | min7 | (1, b3, 5, b7) |
| (Major-Minor) Seventh | 7 | (1, 3, 5, b7) |
| Diminished Seventh | dim7 | (1, b3, b5, bb7) |
| Half Diminished Seventh | hdim7 | (1, b3, b5, b7) |
| Minor (Major Seventh) | minmaj7 | (1, b3, 5, 7) |
| **Sixth Chords:** | | |
| Major Sixth | maj6 | (1, 3, 5, 6) |
| Minor Sixth | min6 | (1, b3, 5, 6) |
| **Extended Chords:** | | |
| Ninth | 9 | (1, 3, 5, b7, 9) |
| Major Ninth | maj9 | (1, 3, 5, 7, 9) |
| Minor Ninth | min9 | (1, b3, 5, b7, b9) |
| **Suspended Chords:** | | |
| Suspended 4th | sus4 | (1, 4, 5) |
| Suspended 2nd | sus2 | (1, 2, 5) |

**Summary Statistics**

The corpus comprises 64 files, one for each theme. The average number of measures per theme is 8.8, with a standard deviation of 5.5 measures. The longest theme in the corpus is 35 measures and the shortest is technically one measure ("Duel of the Fate Ostinato" is encoded as one measure between repeat signs). Of the 64 files, 50 have been given an exclusive key interpretation. This means that approximately 20% of the corpus, or 9 themes, can be considered not to have any prevailing key. This subset of material therefore contains no functional harmony interpretations. Interestingly, only 14 files (themes) are nominally in major mode, with the other 36 files nominally minor. Some general corpus census statistics are presented in Table 3.

**Table 3:** Basic census data for the SWTC

| Number of notes (incl. ties) | 4,008 |
|---|---|
| Number of onsets (ignore ties) | 3,866 |
| Longest note | Dotted whole note |
| Shortest note (excluding grace notes) | 64th note |
| Number of rests | 294 |
| Highest note | F#6 |
| Lowest note | B1 |
| Number of total measures | 542 |

A distribution of the pitch content is illustrated in Figure 2. While the bulk of thematic content can be seen as distributed over the two-octave range above C4 (middle C), this figure illustrates the impressive pitch range of John Williams' thematic material. Figure 3 illustrates the most prevalent chord types in descending order. Finally, Figure 4 shows each file in the corpus–grouped by trilogy in chronological order–illustrating the duration of each file both in terms of overall duration in seconds as well as via note counts. Songs with a larger difference between note counts and duration therefore exhibit greater note density.

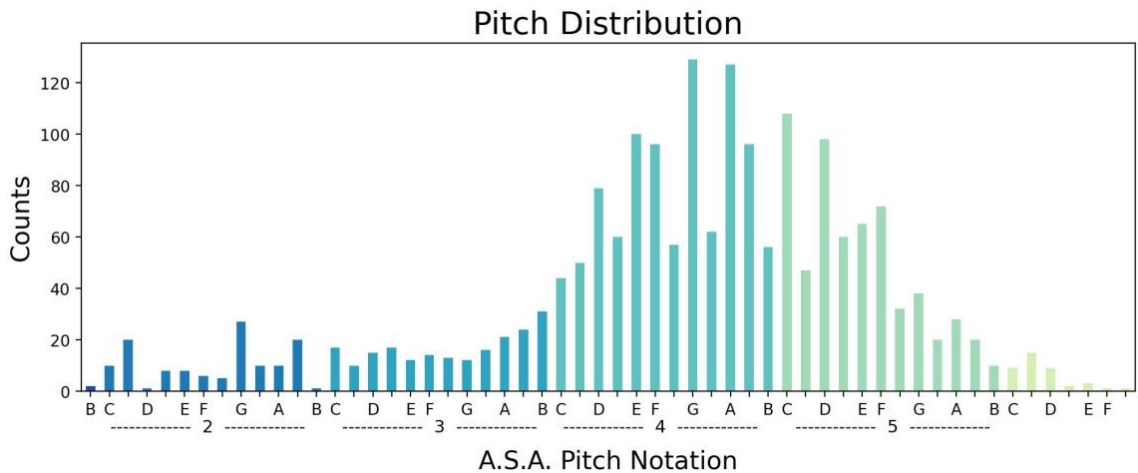

**Figure 2.** Pitch distribution in the STWC. Octave register is indicated below each new octave (e.g., C4-B4, C5-B5, etc.) and each shaded a different color.

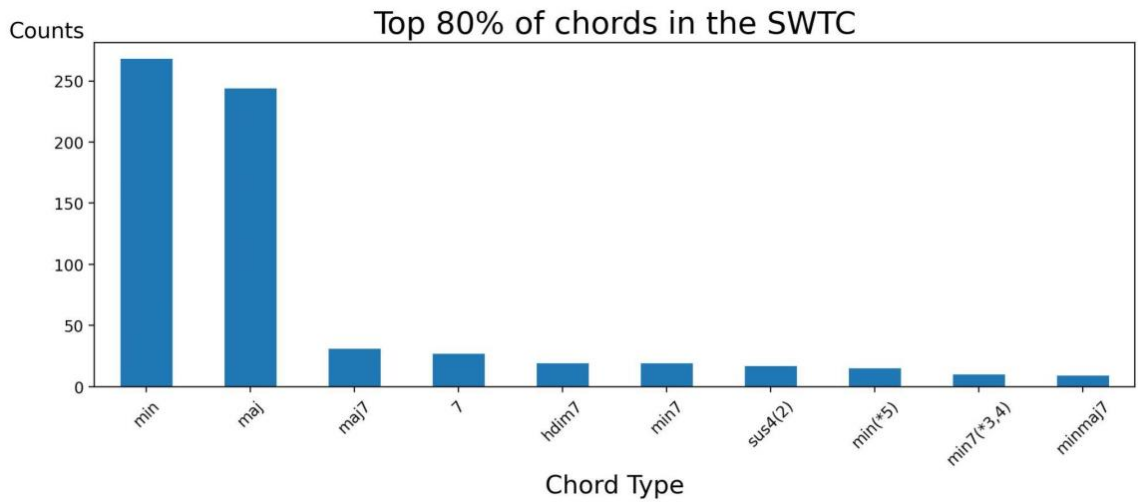

**Figure 3.** Distribution of chord types (top 80% only) in the SWTC. This ignores distinctions based on bass notes (inversions). Note the high prevalence of chord types such as sus chords and minor-major 7th chords which are practically non-existent in music of the Common Practice Era (e.g., see Devaney et al., 2015).

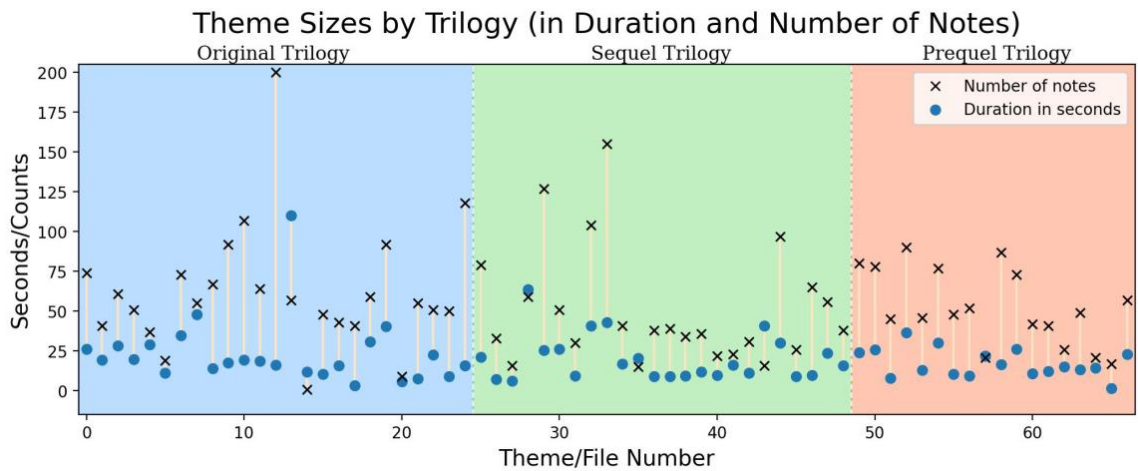

**Figure 4..** Duration of each theme (file) as measured in terms of the number of total notes (x) and the total duration in seconds (●). The three collections of themes belonging to each trilogy are placed in chronological order.

## DISCUSSION

While the SWTC is fairly small for a modern digital corpus, it contains the raw digital scores of the transcriptions, and the symbolic encoding of those transcriptions in humdrum format, including melodic and harmonic information. All of this information was painstakingly curated by multiple experts. Recently, two other corpus analyses of film music (likewise from expert-created custom corpora) have been published (Richards, 2016; Yorgason & Lyon, 2020, also Lyon & Yorgason, 2021), however, only the latter have made the corpus files available for public use. The present work was built specifically for the purpose of facilitating computational analysis. We do aim to make future, significantly expanded releases of this corpus. Future releases would include ancillary themes that do not fit the current definition as a "leitmotif" proper, as well as (potentially) the harmonic information in RomanText format (Tymoczko et al., 2019) for facilitating use with the music21 toolkit (Cuthbert & Ariza, 2010). Author FL has already cataloged over 18 hours of additional musical material that will hopefully be added to the future iterations of the SWTC.

Part of the goal in publicly sharing this corpus is to encourage computational analysis of film music–an understudied topic in music theory. By providing harmony annotations in multiple formats, including a new **harte humdrum standard syntax, we aim to facilitate different types of analyses. We hope that our attention to detail in systematically annotating notoriously difficult musical features (such as non-functional harmonies, polychords, etc.) will push other corpora to include similar "difficult" features. And while *Star Wars* is unique in the complexity and density of its musical materials, many other major franchises (Marvel Cinematic Universe, Lord of the Rings) involve similarly rich thematic networks; we hope for SWTC to serve as a model for future explorations along these lines. Indeed, we hope this corpus may provide the foundation for uncovering aspects of the elusive "film music sound" while also being a useful corpus against which to compare styles from other repertoires such as classical, jazz, and pop.

## NOTES

[1] Correspondence can be addressed to: Claire Arthur, Georgia Institute of Technology, claire.arthur@gatech.edu.